\documentclass[a4]{paper}
\usepackage[cp1251]{inputenc}
\usepackage[T2A]{fontenc}
\usepackage[english]{babel}
\usepackage{amssymb}
\usepackage{amsmath}
\usepackage{textcomp}
\usepackage{indentfirst}
\usepackage{float}
\usepackage{caption}
\usepackage{textcomp}
\DeclareCaptionLabelSeparator{pipe}{ $: $ }
\usepackage{authblk}

\usepackage{units}
\usepackage{amsfonts}
\usepackage{graphicx} 
\usepackage{amstext} 
\usepackage{wasysym} 

\usepackage{comment}
\usepackage{color}

\usepackage{epstopdf}
\usepackage{xcolor}
\usepackage[citecolor=green, linkcolor=red]{hyperref}
 
\definecolor{urlcolor}{HTML}{000000} 

\numberwithin{equation}{section}

\hypersetup{citecolor=green, linkcolor=red,urlcolor=urlcolor}
\usepackage[left=2.5cm,right=2cm, top=2cm,bottom=2cm,bindingoffset=0cm]{geometry} 

\usepackage[sorting=none,backend=bibtex,citestyle=numeric]{biblatex}
\bibliography{literature.bib}
 
\begin{document}

\title{Isotropization by shock waves generation in anisotropic hydrodynamics}

\author{Aleksandr Kovalenko}
\affil{P.N. Lebedev Physical Institute, Moscow, Russia}

\maketitle

\begin{abstract}
Anisotropic hydrodynamics (aHydro) has proven successful in modeling the evolution of quark-gluon matter created in heavy-ion collisions. The hydrodynamic description of quark-gluon plasma has also been widely used to study sound phenomena, such as shock waves. It has recently been shown that initial fluctuations in energy density and supersonic partons can generate fairly strong shock waves. However, significant anisotropic properties of the system due to the rapid longitudinal expansion of matter have not been taken into account in such studies. Moreover, the process of isotropization and its characteristic time-scales were not considered along with the question of the shock waves formation. Previous studies on shock discontinuous solutions in anisotropic hydrodynamics assumed constant anisotropy, leading to flow refraction towards the anisotropy axis and flow acceleration, characteristics of rarefaction waves, indicating limitations in this approach. This paper investigates discontinuous solutions for normal shock waves without flow refraction, introducing two compression parameters for longitudinal and transverse pressures. The resulting analytical solutions, as well as numerical computations, provide an isotropization mechanism of the system.
\end{abstract}

\section*{Introduction}

In recent decades, the idea of using a hydrodynamic description for quark-gluon matter formed in ultrarelativistic collisions of heavy ions took hold \cite{Heinz2005, Heinz2013, Romatschke2017}. The collective expansion of matter leads to such observable phenomena as transverse flow and elliptical flow \cite{Gustafsson1984, NA49:1998}. Moreover, it has been shown that this matter exhibits properties of an ideal fluid. This has led to a great interest in the hydrodynamic description of the evolution of quark-gluon plasma and the corresponding phenomena, in particular shock waves, which have already been studied previously for nuclear matter and ultrarelativistic gas \cite{Scheid1974,Israel, Mitchell, Majorana1990}. It was shown that initial fluctuations of energy density can generate transverse shock waves in quark-gluon matter already at the earliest stages of evolution \cite{Gyulassy:1996ka, Gyulassy:1996br}. Experiments on two-particle azimuthal correlations in the context of the jet quenching effect led to interest in studying the Mach cone \cite{Satarov2005, Betz2007, CasalderreySolana2007, Roy2009}. It was also shown that the trigger jet energy is sufficient to form strong shock waves \cite{Shuryak2012}.

Viscous hydrodynamic theories have proven to be the most successful in describing quark-gluon plasma (QGP) \cite{Muronga:2002, Kolb:2003dz, Baier:2006um, Romatschke:2009im, Calzetta:2015}. Nevertheless, quark-gluon plasma becomes highly anisotropic at the early stages after the collision due to the rapid longitudinal expansion of matter. Already in the first-order Navier-Stokes theory, it is possible to obtain an expression for the ratio of longitudinal and transverse pressures $P_\parallel / P_\perp$ in a case of transversely homogeneous and boost invariant in the longitudinal direction system \cite{Strickland}. For a fixed viscosity $\eta/S$ we have $P_\parallel / P_\perp < 1$ , and this anisotropy increases with decreasing temperature. Such a large momentum-space anisotropy pose a problem for 2nd-order viscous hydrodynamics, since it based on a linearization around an isotropic background. It has been shown that anisotropy leads to large corrections, which in turn lead to unphysical results during evolution, such as negative pressure. A new approach to treat these problems is anisotropic hydrodynamics (aHydro) \cite{MartStr, RybFlor, Strickland, Alqahtani:2017mhy}. In contrast to the standard viscous methods, new anisotropic approach takes into account anisotropy  already in the leading order of the hydrodynamic expansion. Anisotropic hydrodynamics yields solutions much closer to the exact Boltzmann equation than standard viscous frameworks, for longitudinally boost-invariant systems \cite{Florkowski:2013}, as well as Gubser flow \cite{Nopoush2015, Martinez2017}. This approach has also proven promising in modeling experimental data from heavy-ion collision \cite{Mubarak:2017, Alqahtani:2018fcz}.

The strong anisotropic properties of the quark-gluon plasma have raised questions regarding the mechanisms and timescales of isotropization of the system \cite{Kovchegov2005, Epelbaum2013, Strickland2013}. It has been demonstrated that the initial gauge fields fluctuations of the weakly coupled QGP at the early stage of the evolution of quark-gluon matter can accelerate the isotropization process \cite{Akkelin2008}. Hard-loop simulations of chromo-Weibel instability demonstrate rapid plasma thermalization with a Boltzmann distribution, but significant pressure anisotropies continue to exist for at least 5-6 fm/c \cite{Attems2013}. 

The problem of shock wave formation and system isotropization is closely related to the time scales and viscosity $\eta/S$. It is known from the hydrodynamic description of the QGP that anisotropy increases with increasing viscosity \cite{Strickland}, which affects the rate of isotropization of the system. At the same time, the formation of shock waves also takes some time. It was shown that for $\eta/S > 0.2$ the formation time of shock waves becomes longer than the expected lifetime of the quark-gluon plasma \cite{Bouras2009b}. The mechanism of isotropization of the system by generation of shock waves discovered in this paper links together the time scales of isotropization and formation of shocks.

Previous studies have investigated shock discontinuous solutions of compression shock waves in anisotropic hydrodynamics in the case of constant anisotropy, i.e. the same anisotropy in front of and behind the shock wave $\xi = \xi'$ \cite{Kovalenko2022, Kovalenko2023}. In was shown that such an assumption leads to the effects of flow refraction toward the anisotropy axis (the beam propagation axis) and acceleration of the passed flow. The last circumstance is a characteristic of rarefaction waves, which may imply restrictions on this formulation. Moreover, the assumption that the shock wave does not change the flow anisotropy may lead to a loss of information about the mechanisms of isotropization during evolution of matter. A natural way to obtain discontinuous solutions for the case of $\xi' \neq \xi$ is to fix the flow angle so that there is no flow refraction. However, the lack of reformulation of the compression parameter for the shock wave led to the appearance of an upper limit on the anisotropy parameter \cite{Kovalenko2023b}. Assumptions of constant anisotropy or absence of refraction lead to an equal number of equations and unknowns, which allows one to obtain analytical solutions for shock waves. However, intermediate solutions with the presence of flow refraction are expected to exist, where $\xi' \neq \xi$.

This paper investigates discontinuous solutions of shock waves in the absence of flow refraction. To treat the problem of compression shock waves in anisotropic space two compression parameters (for longitudinal and transverse pressures) was introduced. The plan of the paper is the following. The first section provides a brief exposition of anisotropic relativistic hydrodynamics and obtained discontinuous equations for shock waves. The second and third sections are devoted to analytical solutions for longitudinal and transverse directions of the normal to the shock waves. The fourth section presents the results of the numerical solution of the equations in the general case.

\section{Main equations}

In relativistic anisotropic hydrodynamics, one assumes the one-particle distribution function to be of the Romatschke-Strickland form \cite{MartStr,StrRom1,StrRom2}
\begin{equation}
f(x,p) = f_{iso}\Bigg( \frac{\sqrt{p^\mu \Xi_{\mu\nu}(x) p^\nu}}{\Lambda(x)}\Bigg),
\label{Kin1}
\end{equation} 
where $\Lambda(x)$ is a coordinate-dependent temperature-like momentum scale and $\Xi_{\mu\nu}(x)$ is a coordinate-dependent anisotropy tensor. In the original formulation of anisotropic hydrodynamics a single anisotropy parameter was introduced that expresses the difference between longitudinal and transverse pressures. In this case of one-dimensional (longitudinal) anisotropy one can obtain $(p^\mu \Xi_{\mu\nu}p^\nu = \mathbf{p}^2 +\xi(x) p_\parallel^2)$ in the local rest frame (LRF).

Anisotropic hydrodynamics provide the following form for the energy-momentum tensor
\begin{equation}
T^{\mu\nu} = (\varepsilon + P_\perp)U^\mu U^\nu - P_\perp g^{\mu\nu} + (P_\parallel - P_\perp)Z^\mu Z^\nu,
\label{T_true}
\end{equation}
where $P_\parallel$ и $P_\perp$ -- longitudinal and transverse pressures respectively. Velocity four-vector $U^\mu$ and longitudinal four-vector $Z^\mu$ read
\begin{align}
U^\mu &= (u_0 \cosh \vartheta, u_x, u_y, u_0 \sinh \vartheta),
\label{u_aniso}
\\
Z^\mu &= (\sinh \vartheta, 0,0, \cosh \vartheta),
\label{z_aniso}
\end{align}
where $\vartheta$ -- longitudinal rapidity, $u_x, u_y$ -- transverse velocities and $u_0 = \sqrt{1+u_x^2+u_y^2}$.

The important property of one-particle distribution function (\ref{Kin1}) is that it is possible to extract the anisotropy-dependent part from the pressure and the energy density \cite{MartStr}:
\begin{align}
\varepsilon & = R(\xi) \varepsilon_{\textrm{iso}} (\Lambda),
\label{e} \\
P_{\perp, \parallel} &= R_{\perp, \parallel} (\xi) P_{\textrm{iso}} (\Lambda),
\label{pTpL}
\end{align}
where the ansotropic functions $R_\perp(\xi)$ and $R_\parallel(\xi)$ are
\begin{align}
\label{RTL}   
R_\perp(\xi) = \frac{3}{2\xi} \Bigg( \frac{1 + (\xi^2-1)R(\xi)}{1+\xi}\Bigg), \ \ \ \ R_\parallel(\xi) = \frac{3}{\xi} \Bigg( \frac{(\xi+1)R(\xi) -1}{1+\xi}\Bigg),
\end{align}
\begin{align}
R(\xi) = \frac{1}{2} \Bigg( \frac{1}{1+\xi} + \frac{\arctan \sqrt{\xi}}{\sqrt{\xi}}\Bigg). 
\label{R}
\end{align}

We consider the case of a massless gas, for which the following equation of state exists $\varepsilon = 2P_\perp + P_\parallel$. This equation leads to the following relationship between anisotropic functions: $2 R_\perp(\xi) + R_\parallel(\xi) = 3 R(\xi)$.

The shock wave in the leading-order hydrodynamics can be described by a discontinuous solution of the equations of motion. This equations stems from the requirement that components of energy-momentum tensor normal to the discontinuity hypersurface are discontinuous across it while tangential ones remain continuous \cite{Landau, Mitchell}. The energy-momentum conservation then leads to the following matching condition linking downstream and upstream projections on the direction perpendicular to the discontinuity surface:
\begin{equation}
T_{\mu\nu} N^\mu = T^{'}_{\mu\nu} N^\mu,
\label{gap}
\end{equation}
where  \(N^\mu\) - unit vector normal to the discontinuity surface and  \( T_{\mu\nu} \) and \( T^{'}_{\mu\nu} \) correspond to upstream and downstream energy-momentum tensors correspondingly.  

Consider a flow moving with velocity $v$ at an polar angle \(\alpha\) to the direction of the $Oz$ axis. For normal shock waves the components of the normal vector $N^\mu$ we have $N_\mu = (0, \sin\alpha, 0, \cos\alpha)$. For the downstream flow moving with velocity $v'$, it is assumed that there is no refraction, i.e. $\alpha' = \alpha$.

We will consider compression shock waves for which in the isotropic case $P' > P$. In the anisotropic case one assumes that for two different pressures the compression shock waves leads to $P'_\parallel > P_\parallel, P'_\perp > P_\perp$. Therefore, it is necessary to introduce two quantities
\begin{equation}
\sigma_\perp = \frac{P'_\perp}{P_\perp},\ \ \ \sigma_\parallel = \frac{P'_\parallel}{P_\parallel}.
\end{equation}
These quantities represent a reformulation of the compression parameter $\sigma = P'/P$ in isotropic hydrodynamics that characterize the strength of the shock wave in each direction.

It is convenient to relate the quantities $\sigma_\perp, \ \sigma_\parallel$ as follows:
\begin{align}
\sigma_\parallel = k \sigma_\perp,
\label{k_sigma}
\end{align}
where $k > 0$.

From the definition (\ref{k_sigma}), knowing $\xi, \ k, \ \sigma_\perp$, one can find $\xi'$. However, since $\xi' \geqslant 0$, then a restriction on $k$ arises. For $\xi' = 0$ we get
\begin{align}
k = k_\textrm{lim} (\xi) = \frac{R_\perp(\xi)}{R_\parallel(\xi)}.
\label{k_limit}
\end{align}
And we obtain the following range $k \in (0, R_\perp(\xi)/R_\parallel(\xi)]$.

Thus the matching condition (\ref{gap}) lead to the following system of equations
\begin{align}
\Bigg[ \frac{R_1(\xi)}{1 - v^2} +  \frac{R_2(\xi)}{1 - v^2\cos^2 \alpha} \cos^2 \alpha \Bigg] v - \Bigg[ \frac{R_3(\xi, k)}{1 - v'^2} +  \frac{R_4(\xi, k)}{1 - v'^2\cos^2 \alpha} \cos^2 \alpha \Bigg] \sigma_\perp v' & = 0,
\label{xi_eq_1} 
\\
\Bigg[ R_\perp(\xi) - \sigma_\perp R_\perp(\xi) + \frac{R_1(\xi) v^2}{1 - v^2} -  \sigma_\perp \frac{R_3(\xi, k) v'^2}{1 - v'^2}  \Bigg] \sin \alpha & = 0,
\label{xi_eq_2} 
\\
\Bigg[ R_\perp(\xi) - \sigma_\perp R_\perp(\xi) + \frac{R_1(\xi) v^2}{1 - v^2} -  \sigma_\perp \frac{R_3(\xi, k) v'^2}{1 - v'^2}   + \frac{R_2(\xi)}{1 - v^2 \cos^2 \alpha} - \sigma_\perp \frac{R_4(\xi, k)}{1 - v'^2 \cos^2 \alpha} \Bigg] \cos \alpha & = 0,
\label{xi_eq_3} 
\end{align}
where
\begin{align*}
R_1(\xi) &= R_\parallel(\xi) + 3 R_\perp(\xi), \ \ \ R_2(\xi) = R_\parallel(\xi) - R_\perp(\xi). \\
R_3(\xi, k) &= k R_\parallel(\xi) + 3 R_\perp(\xi), \ \ \ R_4(\xi, k) = k R_\parallel(\xi) - R_\perp(\xi).
\end{align*}

One can see from equations (\ref{xi_eq_1} - \ref{xi_eq_3}) that for $\alpha = 0, \pi/2$ only two equations remain due to the multiplication of expressions (\ref{xi_eq_2} - \ref{xi_eq_3}) by trigonometric functions. However, for any other values of $\alpha$ we will consider the expressions in square brackets in equations (\ref{xi_eq_2} - \ref{xi_eq_3}). To preserve the continuity of the solutions, we should neglect $\sin \alpha$ in (\ref{xi_eq_2}) and $\cos \alpha$ in (\ref{xi_eq_3}) also for $\alpha = 0, \pi/2$.

We will solve the equations with respect to the quantities $k, v, v'$. Then from the equations (\ref{xi_eq_2} - \ref{xi_eq_3}) one finds
\begin{equation}
k = \frac{\sigma_\perp R_\perp + R_\parallel - R_\perp - [\sigma_\perp v^2 R_\perp - v'^2 (R_\perp - R_\parallel)] \cos^2 \alpha }{\sigma_\perp (1 - v^2 \cos^2\alpha) R_\parallel}.
\label{k_answer}
\end{equation}

\section{Longitudinal case}

If the normal vector $N^\mu$ is parallel to the anisotropy axis, then such a shock wave will be called longitudinal. In this case we have $v = v_z, \ v' = v'_z$ and $N^\mu = (0, 0,0, 1)$. From the equations (\ref{xi_eq_1}, \ref{k_answer}) one obtains
\begin{equation}
v'_z = \sqrt{\frac{R_\parallel^3 + 4R_\parallel^2 R_\perp + 3R_\parallel R_\perp^2 - 2 R_\perp^3 + (9R_\perp^3 + R_\parallel^2 R_\perp + 8 R_\parallel R_\perp^2) \sigma_\perp - 2R_\perp (2R_\perp + R_\parallel) S(\xi, \sigma_\perp)}{(R_\parallel - R_\perp) (R_\parallel^2 - R_\perp^2(\sigma_\perp - 6) + R_\parallel R_\perp (\sigma_\perp + 5))}},
\label{v2_long}
\end{equation}
where  $R_\parallel = R_\parallel(\xi), \ R_\perp = R_\perp(\xi)$ and
\begin{equation*}
S(\xi, \sigma_\perp) =S(R_\perp(\xi), R_\parallel(\xi), \sigma_\perp)  = \sqrt{4 R_\parallel^2 \sigma_\perp + R_\perp(R_\perp + 4 R_\parallel \sigma_\perp (1 + \sigma_\perp) + R_\perp \sigma_\perp (5\sigma_\perp - 2) )}.
\end{equation*}
For the downstream flow one finds
\begin{equation}
v_z = v'_z \frac{R_\parallel^2 + R_\parallel R_\perp (1+\sigma_\perp) + R_\perp [R_\perp(2 - \sigma_\perp) + S(\xi, \sigma_\perp)]}{(R_\parallel + R_\perp) (R_\parallel + R_\perp (2 + \sigma_\perp))}.
\label{v_long}
\end{equation}

If we expand these expressions for small $\xi$ and then let it tend to zero, we obtain the well-known isotropic solution for massless gas
\begin{equation}
v_z = \sqrt{\frac{1 +  3 \sigma }{3 (3 + \sigma)}}, \, \, v'_z = \sqrt{\frac{3 + \sigma }{3 (1 +  3 \sigma)}}.
\label{v_iso}
\end{equation}
where $\sigma = \sigma_\perp (\xi = 0) = P'/P$ and $P, \ P'$ -- isotropic pressures.

In fact, the value of $k$ for $\xi >0$ separates the solutions for the shock wave into solutions of compression waves ($k > 1, \ \sigma_\perp > 1, \ v' < v$) and solutions of rarefaction shock waves ($k < 1, \ \sigma_\perp < 1, \ v' > v$). For compression shock waves we obtain strict relations $v' < v, \ \xi' < \xi$. Thus, the downstream flow loses speed and becomes isotropized. The opposite situation is for the case $k < 1$, when we obtain $v' > v, \ \xi' > \xi$, so the anisotropy of the flow increases. Since we are interested in compression waves, the case $k < 1$ will not be considered. For $k = 1$ we must consider only the isotropic solution for $\xi = 0$.

It is worth noting that for the solutions (\ref{v2_long} - \ref{v_long}) we do not obtain the entire range of $k$ values. Namely, for compression shock waves we have $k \in (1,k*]$, where $k* < k_\textrm{lim} (\xi) = R_\perp(\xi)/R_\parallel(\xi)$ and
\begin{equation}
k* (\xi) = \frac{-R_\perp + \sqrt{R_\perp (4 R_\parallel + 5 R_\perp)}}{2 R_\parallel}.
\label{k_special}
\end{equation}
The reason for this behavior is clear if we solve the equations (\ref{xi_eq_1} - \ref{xi_eq_3}) for $v, v', \sigma_\perp$, and take $k$ as the input parameter. In this case, we obtain a solution in which $v > 1$ for $k > k*$. On the other hand we have $\sigma_\perp \rightarrow \infty$ when $k \rightarrow k*$.

\begin{figure}[H]
\center{\includegraphics[width=1\linewidth]{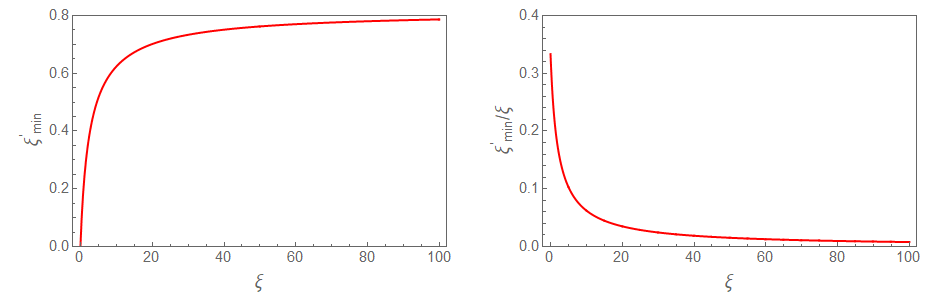}}
\caption{\small  Plot of $\xi'_\textrm{min} (\xi)$ (left) and the ratio $\xi'_\textrm{min} (\xi) / \xi$ (right).}
\label{xi_min_alpha_0}
\end{figure}

In other words, in such a solution there is a constraint on the minimum value $\xi' = \xi'_\textrm{min}$ for a given $\xi$. Fig. \ref{xi_min_alpha_0} shows a graph of the dependence of the minimum value $\xi'_\textrm{min}$ on $\xi$. It is evident that this constraint is significant for small values of $\xi$. The ratio of $\xi'_\textrm{min}$ to $\xi$ tends to $1/3$ for $\xi \rightarrow 0$ and decreases monotonically to zero for $\xi \rightarrow \infty$.

As we will see in the following sections, for any $\alpha > 0$ solutions are possible for all values of $k \in (1, R_\perp(\xi)/R_\parallel(\xi)]$. For small $\alpha$ the transition near $k = k*$ occurs at a some very large value of $\sigma_\perp = \sigma*$ (such that $v \approx 1$). Accordingly, if $\alpha \rightarrow 0$, then we obtain $\sigma* \rightarrow \infty$. Thus, the analytical constraint on $k$ when $\alpha = 0$ are fundamental features of the longitudinal case under consideration.

\begin{figure}[H]
\center{\includegraphics[width=1\linewidth]{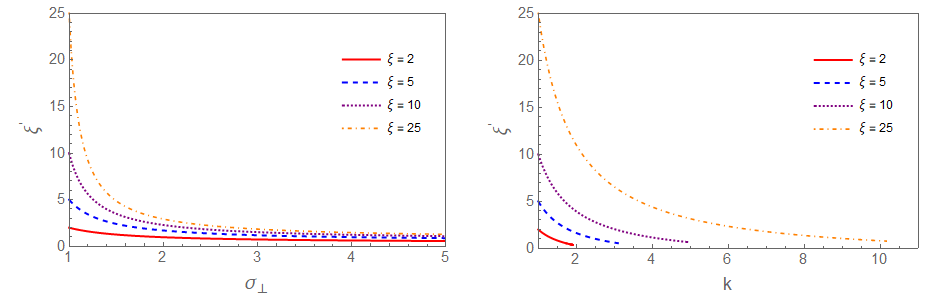}}
\caption{\small Plot of $\xi'$ as a function of $\sigma_\perp$ (right) and as a function of $k$ (left) for a different values of $\xi$ in longitudinal case ($\alpha = 0$).}
\label{xi2_parallel}
\end{figure}

Fig. \ref{xi2_parallel} shows a plot of the dependence of $\xi'(\sigma_\perp), \ \xi'(k)$ for different $\xi$, which has a monotonically decreasing character. Moreover, as noted earlier, the flow is isotropized, i.e. $\xi' < \xi$.

\section{Transverse case}

Consider the case when the normal vector $N^\mu$ is perpendicular to the anisotropy axis (transverse shock wave) and choose the $x$-axis due to the symmetry in $Oxy$ plane. Then we have $v = v_x, \ v' = v'_x$ and $N^\mu = (0, 1, 0, 0)$. From the equations (\ref{xi_eq_1}, \ref{k_answer}) one can obtain the following solutions for the velocities of upstream and downstream flows:
\begin{equation}
v_x = \sqrt{\frac{R_\parallel +  3 \sigma_\perp R_\perp }{3 [R_\parallel +R_\perp (2 + \sigma_\perp) ]}}, \, \, v'_x = \sqrt{\frac{ R_\parallel +R_\perp (2 + \sigma_\perp)}{3 [R_\parallel +  3 \sigma_\perp R_\perp]}}.
\label{v_perp}
\end{equation}
It is important to note that the product of these velocities $vv'$ gives exactly $1/3$, as in the isotropic solution (\ref{v_iso}). In the absence of anisotropy (which also corresponds to $\sigma_\perp = \sigma_\parallel = \sigma)$ these solutions transform into the well-known formulas for the ultrarelativistic isotropic case (\ref{v_iso}).

One can also see how the formulas (\ref{v_perp}) differ from those obtained with constant anisotropy \cite{Kovalenko2022}:
\begin{equation}
\label{velT}
v_x = \sqrt{\frac{R_\perp(3 \sigma R + R_\perp)}{3R(R_\perp\sigma + 3 R))}}, \;\;\; 
v'_x= \sqrt{\frac{R_\perp (R_\perp\sigma + 3 R)}{3R(3 \sigma R + R_\perp)}},
\end{equation}
where $\sigma = P'_{\textrm{iso}}/P_{\textrm{iso}}$ and pressures are defined in (\ref{pTpL}).

Simple analytical equations (\ref{v_perp}) lead to the solution for $k$:
\begin{equation}
k |_{\alpha = \pi/2} = \frac{R_\parallel + R_\perp (\sigma_\perp - 1)}{ \sigma_\perp R_\parallel}.
\label{k_perp}
\end{equation}

\begin{figure}[H]
\center{\includegraphics[width=1\linewidth]{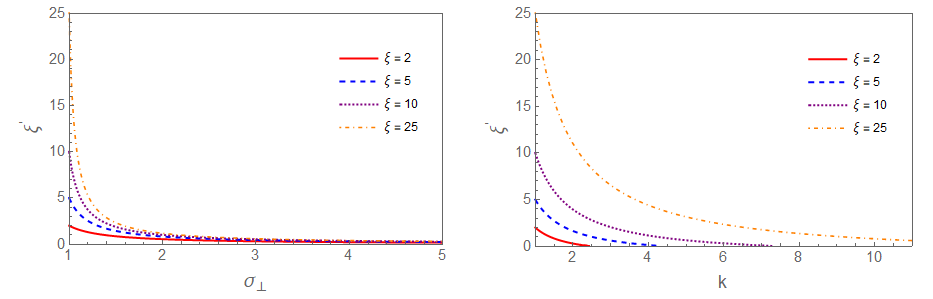}}
\caption{\small Plot of $\xi'$ as a function of $\sigma_\perp$ (right) and as a function of $k$ (left) for a different values of $\xi$ in transverse case ($\alpha = \pi/2$).}
\label{xi2_perp}
\end{figure}

On Fig. \ref{xi2_perp} one can see that the dependence of $\xi'$ in similar to the longitudinal case. The difference is that $\xi'$ decreases more quickly with increasing $\sigma_\perp$.

\section{Case of arbitrary polar angle}

For an arbitrary polar angle $\alpha$, the equations (\ref{xi_eq_1} - \ref{xi_eq_3}) are solved numerically. It is convenient to consider solutions for the velocities $v, \ v'$ as functions of $k$ rather than $\sigma_\perp$, so in this case we do not have to work with insanely large values of $\sigma_\perp$ and we can see a full range of velocities. 

\begin{figure}[H]
\center{\includegraphics[width=1\linewidth]{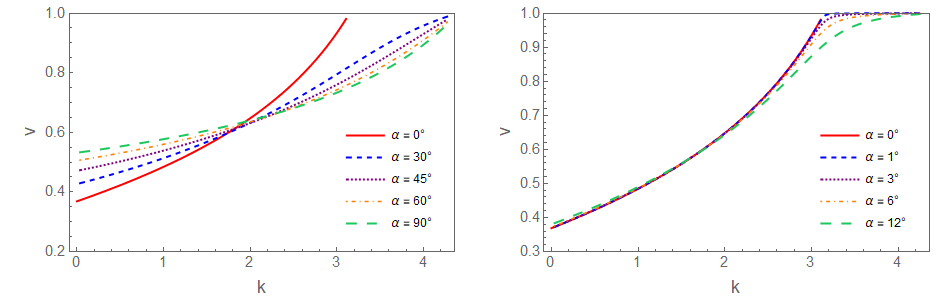}}
\caption{\small Plot of $v$ as a function of $k$ for a different values of $\alpha$ with $\xi = 5$. The case of small angles is shown on the right plot. }
\label{v_alpha}
\end{figure}

Fig. \ref{v_alpha} shows graphs of the dependence of the velocity of the upstream flow $v$ on $k$ for different polar angles $\alpha$. One can see that for small angles the solution bends near the value $k = k*$ and the value $v = 1$ becomes an asymptote.

\begin{figure}[H]
\center{\includegraphics[width=1\linewidth]{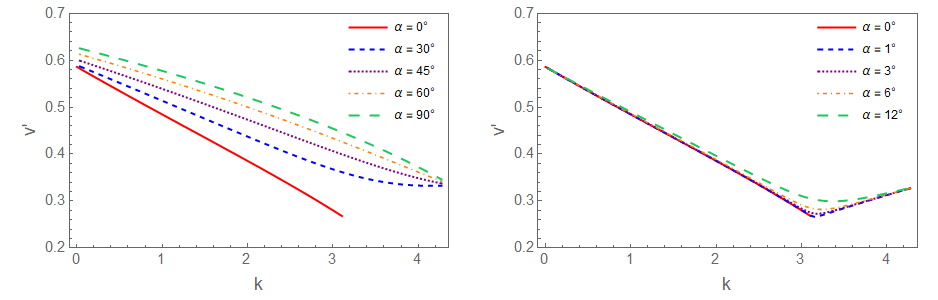}}
\caption{\small Plot of $v'$ as a function of $k$ for a different values of $\alpha$ with $\xi = 5$. The case of small angles is shown on the right plot.  }
\label{v2_alpha}
\end{figure}

Also a plot can be constructed for the velocity of the downstream flow $v'$ (see Fig. \ref{v2_alpha}). For small angles a minimum is formed near the point $k = k*$ (\ref{k_special}) at which $\xi' = \xi'_\textrm{min}$.

\begin{figure}[H]
\center{\includegraphics[width=1\linewidth]{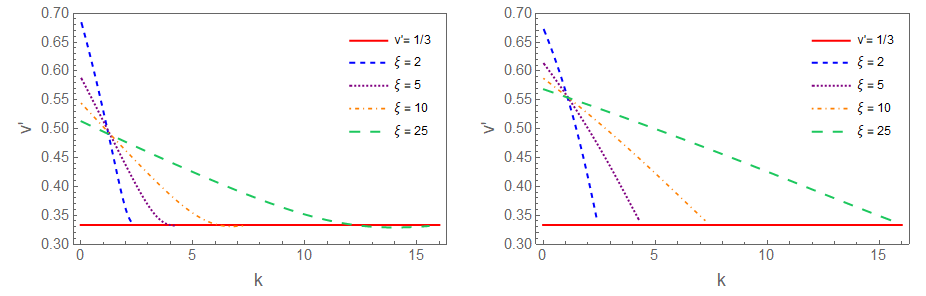}}
\caption{\small Plots of $v'$ as a function of $k$ for a different values of $\xi$ with $\alpha = \pi /6$ (left) and $\alpha = \pi /3$ (right). }
\label{v2_limit}
\end{figure}

In the isotropic ultrarelativistic solution (\ref{v_iso}), as $\sigma \longrightarrow \infty$, one obtains two limits $v = 1$ and $v' = 1/3$. In fact, regardless of the angle $\alpha$ (except for the special case $\alpha = 0$) and $\xi$, as $k$ increases to $k_\textrm{lim}$ (that is, as $\sigma_\perp \rightarrow \infty$), the velocities $v, \ v'$ tend to the values $v \rightarrow 1, \ v' \rightarrow 1/3$, as in the isotropic case. In Fig. \ref{v2_limit}, one can see that the minimum of $v'$ is below the value $v' = 1/3$ for small angles, but in the limit $k \rightarrow k_\textrm{lim}$ one obtains $v' \rightarrow 1/3$.

To characterize the transformation of the flow velocicites $v(\xi, \alpha, \sigma_\perp) \to v'(\xi, \alpha, \sigma_\perp)$ the relative difference between velocities of downstream and upstream flows is introduced:
\begin{equation}
\delta (\xi, \alpha, \sigma_\perp) = \frac{v (\xi, \alpha, \sigma_\perp)' - v(\xi, \alpha, \sigma_\perp)}{v(\xi, \alpha, \sigma_\perp)}.
\end{equation}

\begin{figure}[H]
\center{\includegraphics[width=1\linewidth]{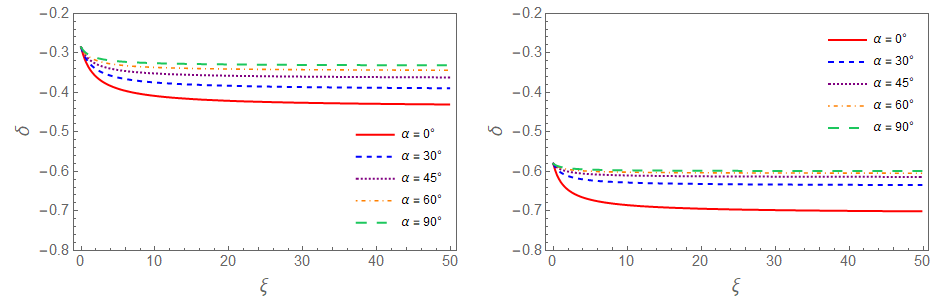}}
\caption{\small Plots of $\delta$ as a function of $\xi$ for a different values of the polar angle $\alpha$ with $\sigma_\perp = 2$ (left) and $\sigma_\perp = 10$ (right). }
\label{comp_xi}
\end{figure}

Fig. \ref{comp_xi} demonstrates the dependence of $\delta (\xi, \alpha, \sigma_\perp)$ on the anisotropy parameter $\xi$ for different values of the angle $\alpha$ and $\sigma_\perp$. One can see that the nature of the dependence is monotonically decreasing (concave), and the value $\delta$ tends to an asymptotic value with increasing anisotropy. The value of the angle $\alpha$ affects only the position of the asymptote, but does not affect the nature of the dependence itself, namely, larger values of the angle $\alpha$ lead to a smaller value of $|\delta|$.
It was previously obtained (see \cite{Kovalenko2022}) that in the case of constant anisotropy for $\alpha \gtrsim \pi/4$ the dependence of $\delta$ on $\xi$ shows a convex and increasing behavior. However, this characteristic is not observed in the current context. Thus, the presence of anisotropy leads to a stronger shock wave in the sense of a growing difference between the flow velocities. It can also be noted that with an increase in $\sigma_\perp$ and at large angles $\alpha$, the impact of anisotropy on $\delta$ diminishes.

\begin{figure}[H]
\center{\includegraphics[width=1\linewidth]{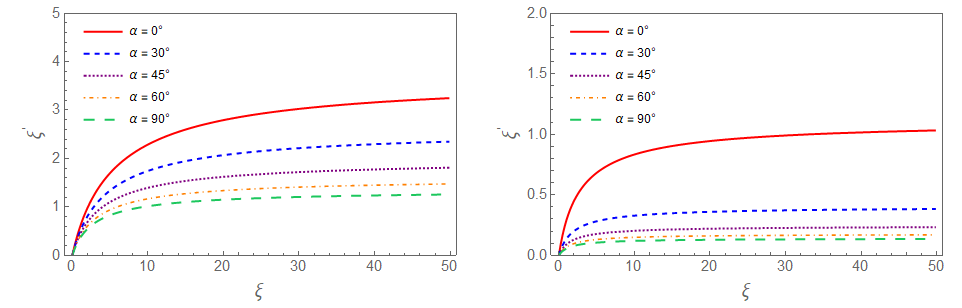}}
\caption{\small Plots of $\xi'$ as a function of $\xi$ for a different values of the polar angle $\alpha$ with $\sigma_\perp = 2$ (left) и $\sigma_\perp = 10$ (right). }
\label{xi2_xi}
\end{figure}

Fig. \ref{xi2_xi} demonstrates the effect of flow isotropization by a shock wave generation. It is evident that the flow is isotropized more strongly with increasing $\alpha$. It can also be noted that the growth of $\xi'$ slows down significantly with increasing $\xi$, but it is not yet clear whether this dependence reaches an asymptote. It can also be said that if shock waves of this type can be generated in QGP, then it introduces a sufficiently strong mechanism for isotropization of the system. High anisotropy in the system cannot be maintained for a long time in system, in which such a shock waves are generated. And, conversely, the formation of shock waves acts as a limiting factor for the increase of anisotropy in the system.

In general, it can be said that near the transverse direction $\alpha \approx \pi/2$ strong shock waves (with large $\sigma_\perp$) almost ignore the presence of anisotropy, so that anisotropy does not affect on relative velocity difference. Moreover, such shock waves strongly isotropize the flow. The relation $vv' = 1/3$ at $\alpha = \pi/2$ also demonstrates similarity with the isotropic description.

Analyzing the right-hand graphs for the small angles in Fig. \ref{v_alpha} and  Fig. \ref{v2_alpha}, one can see that the solutions of the flow velocities $v, \ v'$ approach some limit curves. The case $k < k*$ corresponds to the obtained solutions (\ref{v2_long} - \ref{v_long}). However, in the limit $\alpha \rightarrow 0$ if $k \geqslant k*$  this curve corresponds to $v = 1$ for the upstream flow, and $v' = k R_\parallel (\xi) / (k R_\parallel (\xi) + 2 R_\perp (\xi))$  for the downstream flow. Thus, the full limit curves for $\alpha \rightarrow 0$ are structured as follows:

\begin{equation}
v (\xi, k) |_{\alpha \rightarrow 0} = 
 \begin{cases}
   \frac{\sqrt{R_\perp - R_\parallel} (k R_\parallel + R_\perp)}{\sqrt{(R_\perp - k R_\parallel ) [(k R_\parallel (R_\parallel + 3 R_\perp) + R_\perp (3 R_\parallel + 5 R_\perp)]}} & \text{ if $k < k*$}\\
   1 &\text{if $k \geqslant k*$},\\
 \end{cases}
\end{equation}

\begin{equation}
v' (\xi, k) |_{\alpha \rightarrow 0} = 
 \begin{cases}
   \frac{(R_\perp + R_\parallel) (R_\perp - k R_\parallel)}{\sqrt{(R_\perp - R_\parallel ) [(k R_\parallel (R_\parallel + 3 R_\perp) + R_\perp (3 R_\parallel + 5 R_\perp)]}} & \text{ if $k < k^{*}$}\\
   \frac{k R_\parallel}{ k R_\parallel + 2R_\perp} &\text{if $k \geqslant k^{*}$}\\
 \end{cases}
\end{equation}

\begin{figure}[H]
\center{\includegraphics[width=0.6\linewidth]{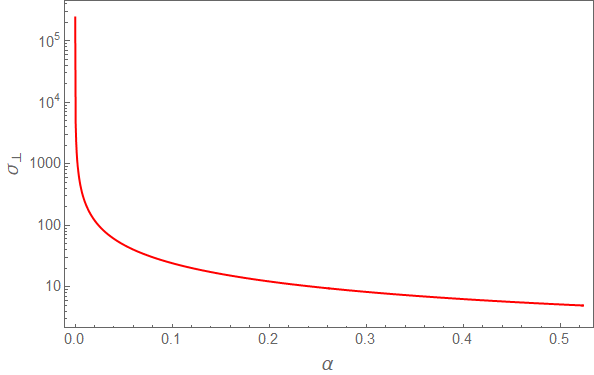}}
\caption{\small Plot of $\sigma_\perp$ as a function of $\alpha$ near the point $k = k^{*}$  with $\xi = 5$. }
\label{s_crit}
\end{figure}

In Fig. \ref{s_crit} it is seen that $\sigma_\perp$ near $k = k^{*}$ reaches very large values as $\alpha$ tends to zero. The plot is given only for the case $\xi = 5$, since its shape depends weakly on $\xi$. More physically acceptable values of $\sigma_\perp$ are reached at $\alpha \gtrsim 0.1$ ($\sim 5^\circ$). 

\section{Isotropization}

In the absence of flow refraction discontinuous solutions correspond to compression shock waves, which demonstrate the mechanism of isotropization of the system. Moreover, the degree of isotropization strongly depends on the direction of the shock wave, i.e. on the angle $\alpha$. For strong shock waves (large $\sigma_\perp$) the flow is almost completely isotropized in the transverse direction. However in the longitudinal direction, a significant amount of anisotropy remains.

\begin{figure}[H]
\center{\includegraphics[width=1\linewidth]{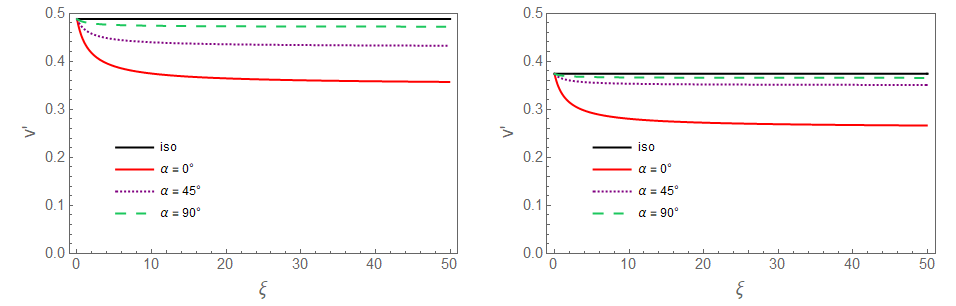}}
\caption{\small Plots of $v'$ as a function of $\xi$ for a different values of the polar angle $\alpha$ with $\sigma_\perp = 2$ (left) and $\sigma_\perp = 10$ (right). Black solid line corresponds to isotropic case }
\label{v2_xi}
\end{figure}

A similar behavior pattern is typical for the difference in flow velocities. For strong shock waves in the transverse direction, the relative difference in flow velocities $\delta (\xi, \alpha, \sigma_\perp)$ differs little from that in the isotropic case and practically does not change with increasing anisotropy (see Fig. \ref{comp_xi}). In the longitudinal direction, the value of $\delta (\xi, \alpha, \sigma_\perp)$ differs significantly from the isotropic case, but a further increase in $\xi$ leads to some asymptotic value. The same behavior is relevant for the downstream flow velocity $v'$, as can be seen from the Fig. \ref{v2_xi}.

In the context of heavy-ion collisions, it is useful to work in standard terms of the transverse momentum $p_T$ and pseudorapidity $\eta$. It is clear that there is an simple transition $v', \alpha \rightarrow p_T, \eta$. In this context, it would be interesting to study the event-by-event fluctuations of the transverse momentum depending on the pseudorapidity. Let us assume that shock waves with behavior similar to the solutions described in this paper are formed in the medium. Then we could consider a correlator of the following form \cite{Acharya2020}:

\begin{equation}
G(\eta_1, \varphi_1, \eta_2, \varphi_2) = \frac{1}{\langle p_{T, 1}\rangle \langle p_{T, 2}\rangle} \Bigg[ \frac{S_{p_T}(\eta_1, \varphi_1, \eta_2, \varphi_2)}{\langle n_{1, 1} (\eta_1, \varphi_1) \rangle \langle n_{1, 2} (\eta_2, \varphi_2) \rangle} - \langle p_{T, 1}\rangle \langle p_{T, 2}\rangle \Bigg],
\end{equation}
with
\begin{equation}
S_{p_T}(\eta_1, \varphi_1, \eta_2, \varphi_2) = \Bigg\langle \sum_{i}^{n_{1, 1}}  \sum_{i \neq j}^{n_{1, 2}}  p_{T, i}  p_{T, j}\Bigg\rangle,
\end{equation}
where $n_{1, 1}$ and $n_{1, 2}$ are the number of tracks on each event within bins centered at $\eta_1, \varphi_1$ and $\eta_2, \varphi_2$. Angle brackets denote averages over an ensemble of events. One can average $G(\eta_1, \varphi_1, \eta_2, \varphi_2)$ across the azimuthal acceptances in which the measurement is performed to obtain $G(\eta_1, \eta_2, \Delta\varphi)$, where $\Delta\varphi = \varphi_2 - \varphi_1$. 

In our case $\eta_1$ is fixed and corresponds to the mid-rapidity region. One might expect that the formation of transverse shock waves would lead to isotropization of the flow in transverse direction  (in the mid-rapidity region). Then the value of $G(\eta_1, \eta_2, \Delta\varphi)$ would reflect the correlations between the isotropized mid-rapidity region and the anisotropic regions. As the difference $\Delta\eta = \eta_2 - \eta_1$ increases, the contribution of anisotropy to the correlator becomes more significant.

\section{Conclusion}

In this paper, analytical solutions for shock waves in the absence of flow refraction were obtained for longitudinal and transverse shock waves. In the longitudinal case, a limitation on the parameter $k$ and, as a consequence, on the anisotropy parameter for the transmitted flow $\xi'$ was shown.

For an arbitrary angle $\alpha$, the common properties for compression shock wave solutions ($k > 1$) are the deceleration of the downstream flow and its isotropization. The degree of isotropization strongly depends on the direction of the shock wave, i.e. on the angle $\alpha$. Properties of strong shocks near the transverse direction ($\alpha \approx \pi/2$) are close to isotropic, i.e. the values of the flow velocities $v, \ v'$ are very close to the known isotropic solutions for a massless gas. Moreover, in the transverse direction, strong isotropization of the system occurs, in contrast to the longitudinal direction. In the limit $\alpha \rightarrow 0$ the solutions for $v, \ v'$ tend to some limit curve for which an analytical expression was obtained. It is important to note that the solutions also demonstrate the possibility of anisotropy enhancement at $k < 1$, but in this case the flow acceleration occurs, which is an attribute of rarefaction shock waves.

To study the obtained isotropization mechanism in the context of heavy-ion collisions, a two-particle correlation function with a fixed mid-rapidity region for the first particle was proposed. This correlation function can be calculated in studies of two-particle event-by-event fluctuations of transverse momentum.

The demonstrated mechanism of isotropization of quark-gluon plasma by shock wave generation links together the questions of isotropization time scales and shock wave formation time scales. The study of this connection is the subject of further research.

\printbibliography

\end{document}